\documentclass[epj]{svjour}
\usepackage{amssymb}
\usepackage{graphics}
\begin{document}
\title{NICA-MPD: azimuthal and femtoscopic particle correlations}

\author{V.A. Okorokov\inst{}
\thanks{e-mail: VAOkorokov@mephi.ru;~Okorokov@bnl.gov}%
}                     
\institute{National Research Nuclear University MEPhI (Moscow
Engineering Physics Institute), \\Kashirskoe Shosse 31, 115409
Moscow, Russian Federation}
\date{Received: 27 September 2015}
%
\abstract{The discussion is focused on the study of the fundamental
symmetries ($\mathcal{P/CP}$) of QCD and geometry of the particle
source. The combination of correlators corresponding to the
absolute asymmetry of distribution of electrically charged
particles with respect to the reaction plane in heavy ion
collisions is studied. A significant decrease of the absolute
asymmetry is observed in the intermediate energy range which can be
considered as indication of possible transition to predominance of
hadronic states over quark-gluon degrees of freedom in the mixed
phase created in heavy ion collisions at intermediate energies.
For the investigation of the energy evolution of the geometric properties
of the particle source the use of femtoscopic radii scaled on the
averaged radius of colliding ions is suggested. This approach
allows the expansion of the set of interaction types, in
particular, on the collisions of non-symmetrical ion beams which can
be studied within the framework of common treatment. There is no
sharp changing of femtoscopic parameter values with increasing of the
initial energy. The suggestions are made for future advancement of
these studies on NICA-MPD.
\PACS{ {25.75.-q} {Relativistic heavy-ion collisions}, {25.75.Gz}
{Particle correlations and fluctuations}
     } 
} 
\maketitle
\section{Introduction}\label{sec:1}

There is a fundamental interrelation between geometry and
fundamental properties of QCD Lagrangian. The vacuum of QCD is a
very complicated matter with rich structure, which can corresponds
to the fractal-like geometry \cite{Okorokov-UJPA-1-1996-2013}. Of
particular interest in the studies of nucleus-nucleus collisions
is a possibility to create a new state of strongly interacting
matter where some of the fundamental symmetries may be violated.
The non-trivial topology of QCD vacuum opens the possibility for
existence of metastable domains which possess of various
properties with respect to the discrete $\mathcal{P/CP}$
symmetries. According to predictions of the theory at finite
temperature \cite{Lee-PRD-8-1226-1973} decays of such domains or
classical transitions (sphalerons) between them in deconfinement
phase of color charges with restored chiral symmetry can result in
local topologically induced violation of $\mathcal{P/CP}$
invariance in strong interactions -- $l\mathcal{TIP}$ effect. In
the presence of background Abelian electromagnetic field the
$l\mathcal{TIP}$ effect can leads to the separation of secondary
charged particles toward the magnetic field
\cite{Kharzeev-NPA-803-227-2008,Fukushima-PRD-78-074033-2008}.
This phenomenon called also chiral magnetic effect (CME) is the
experimental manifestation of the local topologically induced
$\mathcal{P/CP}$ parity violation in strong interactions. On
the other hand, at present femtoscopic measurements in particular
that based on Bose\,--\,Einstein correlations are a unique
experimental method for the determination of the sizes and lifetimes
of the sources, i.e. 4D geometry, in high energy and nuclear physics.
The study of nucleus-nucleus ($A+A$) collisions in wide energy
domain by correlation femtoscopy seems important for better
understanding both the equation of state (EOS) of strongly
interacting matter and the general dynamic features of soft processes.
Therefore, the future experimental studies of the non-trivial
structure of the QCD vacuum and geometry of particle source in the
NICA-MPD energy domain allow an important progress with respect to the
fundamental symmetries and non-perturbative properties of strong interactions.

\section{Fundamental symmetries of QCD at finite temperature}\label{sec:1}

With taking into account possible local strong $\mathcal{P/CP}$
violation the invariant distribution of final state particles with
certain sign of electric charge $\alpha$ ($\alpha = +,-$) can be
written as follows:
$$
E\frac{\textstyle d^{3}N_{\alpha}}{\textstyle d\vec{p}}
=\frac{\textstyle 1}{\textstyle 2\pi} \frac{\textstyle
d^{\,2}N_{\alpha}}{\textstyle
p_{\footnotesize{\perp}}dp_{\footnotesize{\perp}}dy}\biggl[1+2\sum\limits_{n=1}^{\infty}
\sum\limits_{m=1}^{2}k^{m}_{n,\alpha}F^{m}(n \Delta \phi)\biggr].
$$
Here $\Delta \phi \equiv \phi-\Psi_{RP}$, $\phi$ is the azimuthal
angle of the particle under study, $\Psi_{RP}$ the azimuthal angle of the
reaction plane, $F^{1,2}(x) \equiv \cos(x), \sin(x)$,
$k^{1}_{n,\alpha} \equiv v_{n,\alpha}$ is the collective flow of the
$n$-th order, the parameters $k^{2}_{n,\alpha} \equiv
a_{n,\alpha}$ describe the effect of $\mathcal{P/CP}$ violation.
It should be emphasized that the CME is the collective effect and
its investigation is possible via correlation analysis only
because $\langle a_{n,\alpha}\rangle =0 $ if the one-particle
distributions are averaged over the event sample. According to the
theory, the correlator which contains only the contribution of possible
$\mathcal{P/CP}$ violation effect is given by equation
$\langle \mathbf{K}_{n,\alpha\beta}^{T}\rangle = \langle
a_{n,\alpha}a_{n,\beta}\rangle$, where $\alpha, \beta$ are the
electric charge signs of secondary particles. Only the first
harmonic coefficient is analyzed so far because as expected
$a_{1,\alpha}$ accounts for most of the effect under study. Thus
the following notation $\langle
\mathbf{K}_{1,\alpha\beta}^{T}\rangle \equiv \langle
\mathbf{K}_{\alpha\beta}^{T}\rangle$ is used below. The structure
of $\langle \mathbf{K}_{\alpha\beta}^{T}\rangle$ is described
in detail elsewhere
\cite{Kharzeev-NPA-803-227-2008,Okorokov-IJMPE-22-1350041-2013}.
The experimental correlator proposed in
\cite{Voloshin-PRC-70-057901-2004} is defined as follows:
$\langle \mathbf{K}_{\alpha\beta}^{E}\rangle = \langle
\cos\left(\phi_{\alpha}+\phi_{\beta}-2\Psi_{RP}\right)\rangle$.
This observable is sensitive to the effect of possible local
strong $\mathcal{P/CP}$ violation and measures the charge
separation with respect to the reaction plane. Both theoretical
and experimental correlators are averaged over pairs of particles
under study in the event and over all events in the sample. It should be
mentioned that the correlators defined above are $\mathcal{P}$-even
quantities therefore $\langle
\mathbf{K}_{\alpha\beta}^{E}\rangle$ may contain contributions
from background effects unrelated to possible local strong
$\mathcal{P/CP}$ violation. Theoretical and experimental
correlators are related by the following equation:
\begin{equation}
\langle \mathbf{K}_{\alpha\beta}^{E}\rangle = \tilde{B}-\langle
\mathbf{K}_{\alpha\beta}^{T}\rangle, \label{eq:1}
\end{equation}
where $\tilde{B}$ is the total background contribution discussed in
\cite{Okorokov-IJMPE-22-1350041-2013}. The absolute asymmetry with
respect to the reaction plane for the azimuthal distribution of the
electric charges in final state is define as follows
\cite{Okorokov-IJMPE-22-1350041-2013}:
\begin{equation}
A_{a}=-[\langle \mathbf{K}_{\pm\pm}^{E}\rangle-\langle
\mathbf{K}_{\pm\mp}^{E}\rangle]. \label{eq:2}
\end{equation}
In the framework of CME model \cite{Kharzeev-NPA-803-227-2008} for the
ideal case of chiral limit and an extremely large magnetic field
the following relations have been derived
\cite{Okorokov-IJMPE-22-1350041-2013}
\begin{equation}
\langle Q^{2}\rangle \propto \Gamma_{CS},~~~A_{a} \propto
\Gamma_{CS}. \label{eq:3}
\end{equation}
Here $\langle Q^{2}\rangle$ is the average square of the charge
difference (in units of $e$) between opposite sides of the
reaction plane, $\Gamma_{CS}$ is the classical transition rate,
i.e. the Chern--Simons diffusion rate. It should be noted that the 
separation of a possible CME signal from background effects is an
important and difficult task for $\langle
\mathbf{K}_{\alpha\beta}^{E}\rangle$ and consequently for quantity
$A_{{\footnotesize{a}}}$. Based on the available
experimental results and its interpretation for wide initial
energy range one can assume at qualitative level only that the
sizable contribution in $A_{{\footnotesize{a}}}$ will be due to
correlations driven by local $\mathcal{P/CP}$ violation in strong
interactions \cite{Okorokov-IJMPE-22-1350041-2013}.

\begin{figure*}
\vspace*{1cm}
\begin{center}
\resizebox{0.59\textwidth}{!}{%
\includegraphics{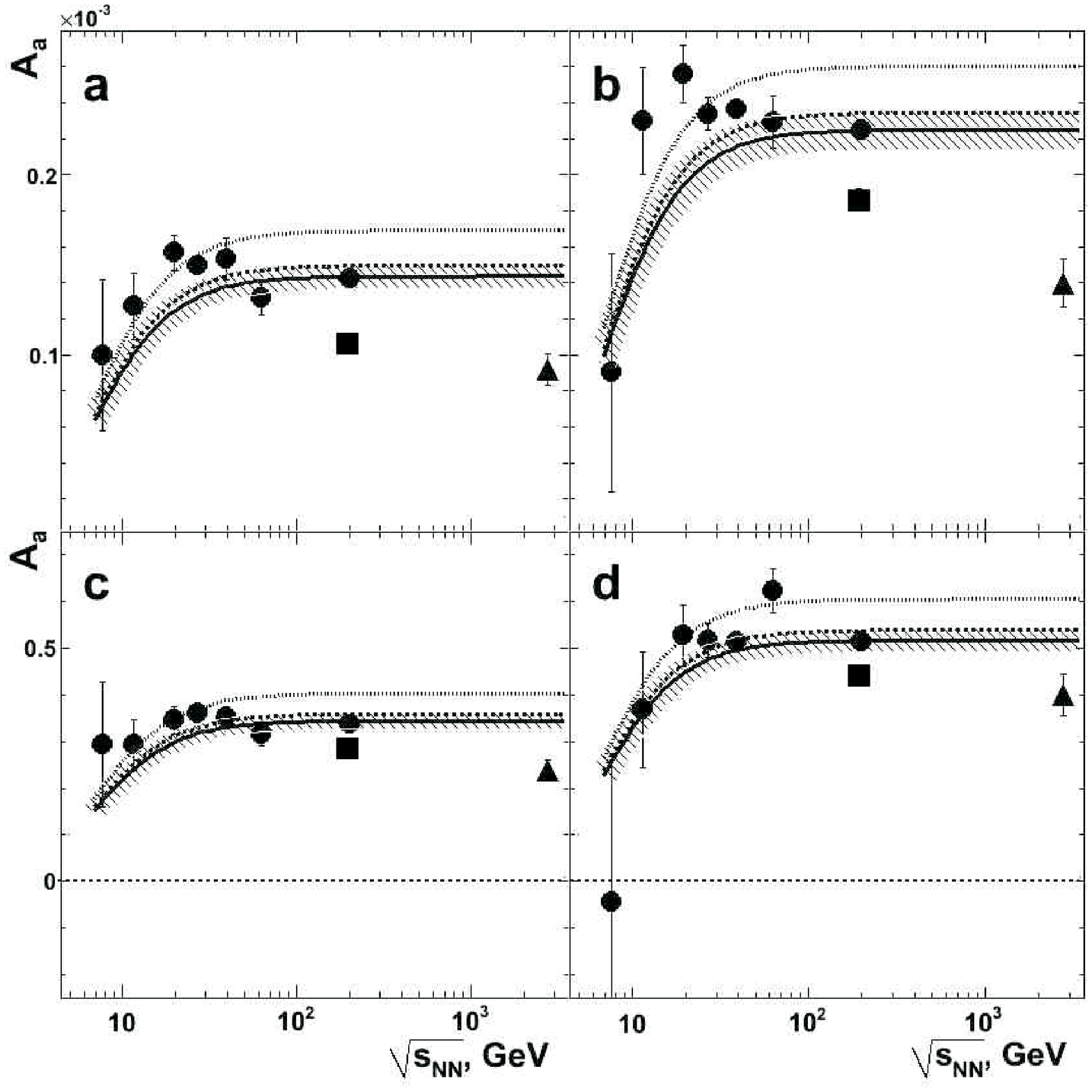}}
\end{center}
\caption{The energy dependence of $A_{a}$ for heavy ion collisions in
various centrality bins: a -- 20-30\%, b -- 30-40\%, c -- 40-50\%
and d -- 50-60\% \cite{Okorokov-IJMPE-22-1350041-2013}.
Experimental points are shown as follows: {\large$\bullet$} --
$\mbox{Au+Au}$, $\blacktriangle$ -- $\mbox{Pb+Pb}$ and
$\blacksquare$ -- $\mbox{U+U}$. Smooth curves correspond to the
energy dependence of the Chern--Simons diffusion rate in the strong
coupling regime for external magnetic field with strength $B=0$
(solid), $B=5T^{2}$ (dashed) and $B=10T^{2}$ (dotted). The shaded
areas for $\Gamma_{CS}$ at $B=0$ are defined by uncertainties of the
$T$ value at fixed initial energy due to errors of the parameters in the
analytic function describing of the experimental dependence $T(\sqrt{s_{NN}})$.} \label{fig:1}
\end{figure*}

Fig. \ref{fig:1} shows the energy dependence of $A_{a}$ for
semi-central events in various bins of centrality for heavy ion
beams. The equation for the Chern--Simons diffusion rate in the case of
finite $B$ had been derived in \cite{Basar-arXiv-1202.2161-2012}. The background magnetic field typically created in relativistic heavy
ion collisions is characterized by the strength $B \sim T^{2}$
\cite{Kharzeev-NPA-803-227-2008,Okorokov-arXiv-0908.2522-2009,Skokov-IJMPA-24-5925-2009}.
The regime of finite external magnetic field is investigated in
the paper and it is observed that the curves for $B=0$ and for the
case $B=T^{2}$ coincide completely for all centrality bins under
study, i.e. for the full centrality domain 20-60\% shown in Fig.
\ref{fig:1}. But relations in (\ref{eq:3}) have been obtained for
extremely strong magnetic field. Thus the energy dependence for
Chern--Simons diffusion rate is computed for external Abelian
magnetic field with various strengths as follows
\cite{Basar-arXiv-1202.2161-2012}:
$$
\Gamma_{CS}^{B \ne 0}=\Gamma_{CS}^{B=0}[1+\zeta^{2} /
(6\pi^{4})],~~~ \zeta \equiv B/T^{2}.
$$
It should be noted that at fixed $\zeta$ the functional dependence of
$\Gamma_{CS}$ on $T$ is the same both for $B=0$ and for the general
case of the presence of a finite external Abelian magnetic field. The
function suggested in \cite{Cleymans-PRC-73-034905-2006} for the
description of the energy dependence of the chemical freeze-out
temperature  agrees with available experimental data quite
reasonably for energies up to $\sqrt{s_{NN}}=200$ GeV
\cite{Aggarwal-PRC-83-024901-2011} and for all centralities
\cite{Kumar-NPA-904-905-256c-2013} under study. Therefore the
analytic dependence $T(\sqrt{s_{NN}})$ from
\cite{Cleymans-PRC-73-034905-2006} is used for estimations of
$\Gamma_{CS}$ in the energy domain under study. The smooth curves in
Fig. \ref{fig:1} are the energy dependencies of the Chern--Simons
diffusion rate in the strong coupling regime at $B=0$ (solid),
$B=5T^{2}$ (dashed) and $B=10T^{2}$ (dotted). The norm for solid
curves is the STAR point for $\mbox{Au+Au}$ at $\sqrt{s_{NN}}=200$
GeV. The shaded bands for the Chern--Simons diffusion rate at $B=0$
are defined by uncertainties of $T$ value at fixed initial energy
due to errors of parameters in the analytical function describing of
$T(\sqrt{s_{NN}})$ experimental dependence
\cite{Cleymans-PRC-73-034905-2006}.
As seen, the phenomenological curves $\Gamma_{CS}(\sqrt{s_{NN}})$ are
in an area caused by the spread of $T$ values at fixed $\sqrt{s_{NN}}$
for following wide range of changing of strength of external
Abelian magnetic field $B \leq 5T^{2}$ at any initial energies
under study, and also at $\sqrt{s_{NN}} \lesssim 12$ GeV for
all the range of $B$ considered in Fig. \ref{fig:1}.
The functional behavior of $\Gamma_{CS}(\sqrt{s_{NN}})$ does not
depend on $B$ in high energy domain at changing of $B$ within the range from $B=0$ until at least the value
$B=10T^{2}$ (Fig. \ref{fig:1}).

One sees the some decreasing of $A_{a}$ for LHC energy as compared
to initial energy domain $\sqrt{s_{NN}} \sim 100$ GeV, the
absolute value of decreasing increases for more peripheral events.
The parameter $A_{a}$ goes down significantly with $\sqrt{s_{NN}}$
decreasing for the intermediate initial energy domain 7.7 -- 20 GeV in
semicentral heavy ion collisions (Fig. \ref{fig:1}). For each
centrality bin under study the estimations for
$A_{a}(\sqrt{s_{NN}})$ correspond to the energy dependence of
$\Gamma_{CS}$ at $B=0$ on the qualitative level with some enhancement
of experimental points over phenomenological curves in the range
$\sqrt{s_{NN}} \sim 20-40$ GeV for events with 20-30\% (Fig.
\ref{fig:1}a) and 30-40\% (Fig. \ref{eq:1}b) centrality. Perhaps,
for better description of $A_{a}(\sqrt{s_{NN}})$ at $\sqrt{s_{NN}}
\sim 20-40$ GeV the influence of external $B$ on $\Gamma_{CS}$
should be taken into account. This suggestion agrees with the
dependence of $B$ on initial energy
\cite{Okorokov-arXiv-0908.2522-2009} which demonstrates that $B$
in the intermediate energy range $\sqrt{s_{NN}} \sim 20-40$ GeV
reaches very large values with respect to the strength of $B$ at
$\sqrt{s_{NN}} \sim 100$ GeV. A qualitative agreement is
observed between values of the $A_{a}$ parameter calculated for
$\mbox{Au+Au}$ collisions and phenomenological curves for any
strength values of $B$ under study. This feature can be considered as
some evidence of validity of relations in (\ref{eq:3}) derived at the
ultimate strong external Abelian magnetic field for heavy ion
collisions. The decrease of $\Gamma_{CS}(\sqrt{s_{NN}})$ observed at
$\sqrt{s_{NN}} < 20$ GeV should lead to attenuation of CME and
its manifestation on experiment at intermediate energies. Taking
into account conditions which are essential for CME one can
suppose the following hypothesis. The change of behavior of the
$A_{a}(\sqrt{s_{NN}})$ dependence observed at the transition from the high
energy domain down to the intermediate energy range may be driven by the
predominance of hadronic colorless states over the quark-gluon
deconfinement phase at $\sqrt{s_{NN}} < 19.6$ GeV and, as a
consequence, by the decrease of CME. Thus the behavior of the experimental
quantity $A_{a}$ vs collision energy agrees with qualitative
expectation for transition to the predominance of the hadronic phase
in the domain $\sqrt{s_{NN}} < 19.6$ GeV and with the decreasing of the
Chern--Simons diffusion rate at intermediate initial energies.

\section{Femtoscopic correlations} \label{sec:3}

The discussion below is focused on the specific case of femtoscopy,
namely, on correlations in pairs of identical char\-ged pions with
small relative momenta -- HBT-in\-ter\-fe\-ro\-met\-ry -- in $A+A$
collisions. The set of main femtoscopic observables $\mathcal{G}
\equiv \{\mathcal{G}_{1}^{i}\}_{i=1}^{4}=\{\lambda,
R_{\mbox{\scriptsize{s}}}, R_{\mbox{\scriptsize{o}}},
R_{\mbox{\scriptsize{l}}}\}$ is only under consideration here. $\mathcal{G}$ characterizes the correlation strength and source's
4-dimensional geometry at the freeze-out stage completely, some
important additional observables which can be calculated with help
of HBT radii are discussed elsewhere
\cite{Okorokov-arXiv-1312.4269,Okorokov-arXiv-1504.08336}. The
most central collisions are usually used for studying the space-time
characteristics of final-state matter, in particular, for the
discussion of the global energy dependence of femtoscopic observables.
Therefore the scaled parameters $\mathcal{G}^{i}$, $i=2-4$ are
calculated as follows \cite{Okorokov-arXiv-1312.4269}:
\begin{equation}
R_{i}^{n}=R_{i}/R_{A},~ i=s, o, l.\label{eq:4}
\end{equation}
Here $R_{A}=R_{0}A^{1/3}$ is the radius of the spherically-symmetric
nucleus, $R_{0}=(1.25 \pm 0.05)$ fm \cite{Valentin-book-1982}. The
change $R_{A} \to \langle R_{A}\rangle=0.5(R_{A_{1}}+R_{A_{2}})$
is made in relation (\ref{eq:4}) in the case of non-symmetric
nuclear collisions \cite{Okorokov-arXiv-1312.4269}. In the general
case the scale factor in (\ref{eq:4}) should take into account
the centrality of nucleus-nucleus collisions. The normalization
procedure suggested in \cite{Okorokov-arXiv-1312.4269} allows the
general consideration of all available data for nucleus-nucleus
collisions together with the proton-proton ($pp$) results at high
energies with replacing $R_{A} \to R_{p}$ in (\ref{eq:4}).
\begin{figure*}
\vspace*{1cm}
\begin{center}
\resizebox{0.59\textwidth}{!}{%
\includegraphics{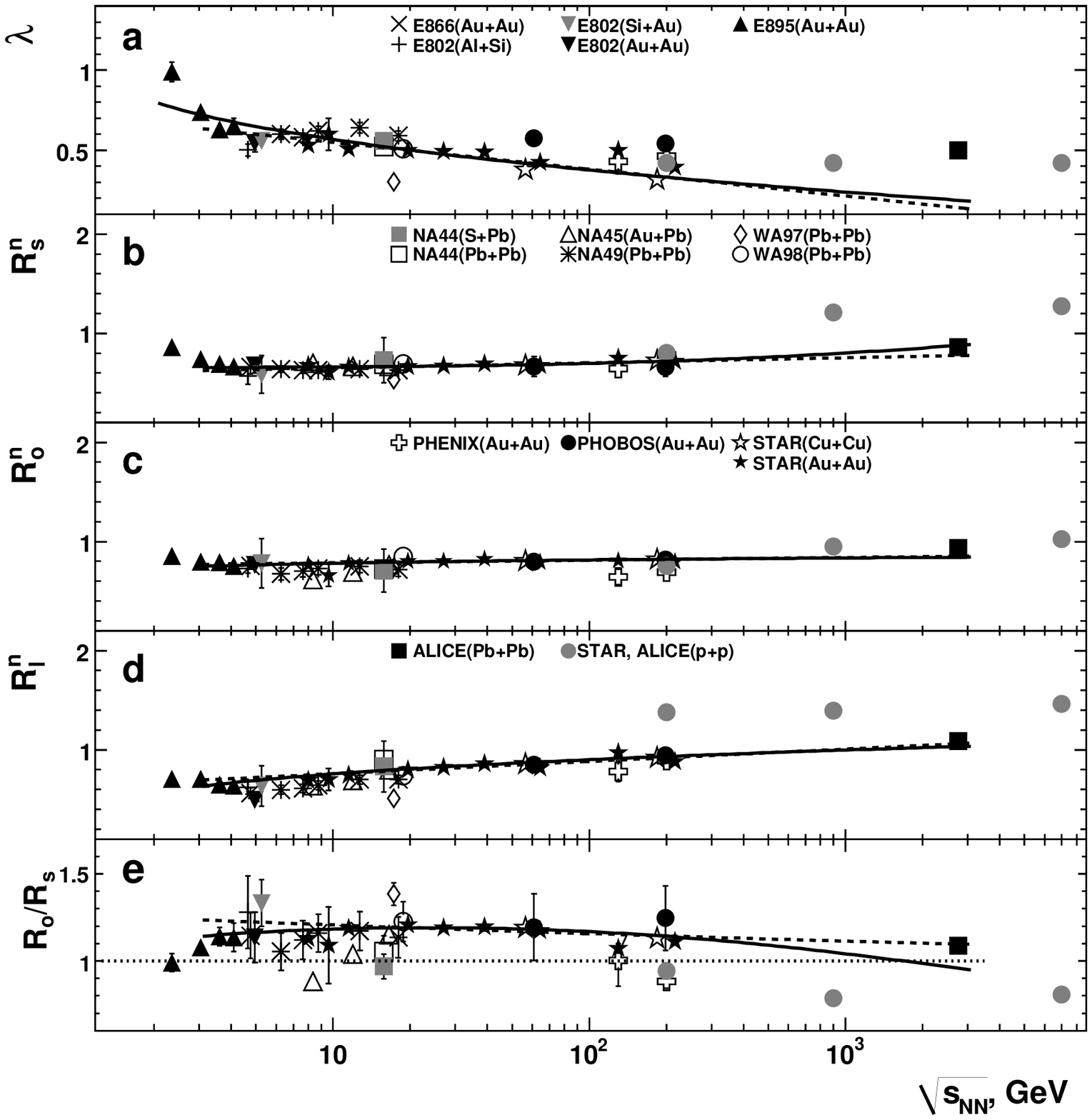}}
\end{center}
\caption{The energy dependence of the $\lambda$ parameter (a), the scaled
HBT-radii (b -- d) and the ratio
$R_{\mbox{\scriptsize{o}}}/R_{\mbox{\scriptsize{s}}}$ (e) in
various collisions. Experimental data are from
\cite{Okorokov-arXiv-1504.08336}. Statistical errors are shown
(for NA44 -- total uncertainties). The solid lines in (a -- d)
correspond to the fits by function (\ref{eq:5}) and the dashed lines
-- to the fits by the specific case of (\ref{eq:5}) at fixed
$a_{3}=1.0$. Smooth solid and dashed curves in (e) correspond to
the ratio $R_{\mbox{\scriptsize{o}}}/R_{\mbox{\scriptsize{s}}}$
calculated from the fit results for
$R^{n}_{\mbox{\scriptsize{s}}}$ and
$R^{n}_{\mbox{\scriptsize{o}}}$ in $A+A$, the dotted line is the level
$R_{\mbox{\scriptsize{o}}}/R_{\mbox{\scriptsize{s}}}=1$.}
\label{fig:2}
\end{figure*}
A detailed study for (quasi)symmetric heavy ion collisions
\cite{Okorokov-arXiv-1312.4269} demonstrates that the fit function
($\varepsilon \equiv s_{\footnotesize{NN}}/s_{0}$, $s_{0}=1$
GeV$^{2}$)
\begin{equation}
f(\sqrt{s_{NN}}) = a_{1}\bigl[1 +
a_{2}(\ln\varepsilon)^{a_{3}}\bigr] \label{eq:5}
\end{equation}
agrees reasonably with experimental
$\mathcal{G}^{i}(\sqrt{s_{NN}})$, $i=1-4$ at any collision energy
for $\lambda$ and at $\sqrt{s_{NN}} \geq 5$ GeV for HBT radii.
Fig. \ref{fig:2} shows the energy dependence of $\lambda$ (a), the
scaled HBT-radii (b -- d) and the $R_{o}/R_{s}$ ratio (e) for both the
symmetric and non-symmetric collisions of various nuclei. Fits of
experimental dependencies for $A+A$ interactions are made by
(\ref{eq:5}) in the same energy domains as well as for
(quasi)symmetric heavy ion collisions. Fit curves are shown in
Fig. \ref{fig:2} by solid lines for (\ref{eq:5}) and by dashed
lines for the specific case of fit function at $a_{3}=1.0$ with taking
into account statistical errors. The fit by (\ref{eq:5})
underestimates the $\lambda$ value at the LHC energy
$\sqrt{s_{NN}}=2.76$ TeV significantly. The $\lambda$ values for
asymmetric nucleus-nucleus collisions at intermediate energies
$\sqrt{s_{NN}} \lesssim 20$ GeV agree well with values of
$\lambda$ in symmetric heavy ion collisions at close energies. On
the other hand the development of some approach is required in
order to account for the type of colliding beams in the case of
$\lambda$ parameter and improve the quality of approximation. Smooth
curves for the normalized HBT radii and the ratio $R_{o}/R_{s}$ are in
reasonable agreement with experimental dependencies in the fitted
domain of collision energies $\sqrt{s_{NN}} \geq 5$ GeV (Figs.
\ref{fig:2}b -- e). The scaled HBT-radii in $pp$ are significantly larger
than those in $A+A$ collisions at close energies.
Because of the feature of Regge theory \cite{Collins-book-1977} the
following relation is suggested to take into account the expansion
of proton with energy:
$R_{p}=r_{0}(1+k\sqrt{\alpha'_{\cal{P}}\ln\varepsilon})$, where
$r_{0}=(0.877 \pm 0.005)$ fm is the proton's charge radius
\cite{PDG-2014}, parameter $\alpha'_{\cal{P}} \propto
\ln\varepsilon$ because the diffraction cone shrinkage speeds up
with collision energy in elastic $pp$ scattering
\cite{Okorokov-AHEP-2015-914170-2015}. $k$ is defined by the
boundary condition $R_{p} \to 1/m_{\pi}$ at $\varepsilon \to
\infty$ with appropriate choice of asymptotic energy
$\sqrt{s_{NN}^{a}}$. A detailed study demonstrates that the
increasing of $\sqrt{s_{NN}^{a}}$ from 6 PeV
\cite{Bourrely-arXiv-1202.3611} to $10^{3}$ PeV influences weakly $R^{n}_{i}$, $i=s, o, l$ in $pp$ collisions and calculations
are made for the first case. The normalized transverse radii agree
in both the $pp$ and the $A+A$ collisions (Figs. \ref{fig:2}b, c)
at $\sqrt{s_{NN}}=200$ GeV with an excess of $R^{n}_{s}$ in $pp$ with
respect to $A+A$ in the TeV-region. The $R^{n}_{l}$ in $pp$ is
larger than that for $A+A$ in the domain $\sqrt{s_{NN}} \geq 200$ GeV.
It should be stressed that additional study is important, at
least, for the choice of $R_{p}(\varepsilon)$.

\section{Future investigations} \label{sec:4}

It should be noted that the requirements for main tracker of MPD
-- time-projection chamber -- give the possibility for
measurements of various correlations, in particular discussed
above, with high quality and statistic.

The study at NICA-MPD with heavy ion beams allows the precise
measurement of the $A_{a}(\sqrt{s_{NN}})$ dependence in the energy
region of sharp decrease of the absolute asymmetry. These future
investigations will be useful for testing the hypothesis about the
attenuation of the CME due to the transition to the predominance of the
hadronic phase in the domain $\sqrt{s_{NN}} < 19.6$ GeV. Furthermore
in this paper it is suggested that at finite temperature ($T$) the
transition rate between vacuum states with different $N_{CS}$ is
dominated by the transition rate due to sphalerons only, without
consideration of the exponentially suppressed transitions due to
instantons. But it should be mentioned that some additional study
and justification may be required for this approach, especially in the
temperature range $T_{c} < T < 3T_{c}$
\cite{Ilgenfritz-PLB-325-263-1994}, where $T_{c}$ is the temperature
of phase transition to the deconfinement state of color charges.
Such investigations can be proposed for the NICA-MPD physics
program.

New experimental data are important for the verification of the
suggestion of separate dependencies $\lambda(\sqrt{s_{NN}})$ for
moderate and heavy ion collisions. The energy dependence is almost
flat for the scaled difference $\delta^{n} \equiv
(R^{2}_{o}-R^{2}_{s})R_{A}^{-2}$ in $A+A$ collisions within large
error bars
\cite{Okorokov-arXiv-1312.4269,Okorokov-arXiv-1504.08336}. The
indication of a possible curve knee at $\sqrt{s_{NN}} \sim 10-20$
GeV obtained in the STAR high-statistics data agrees with other
results in the framework of the phase-I of the BES program at
RHIC. But additional precise measurements at NICA-MPD with various
beams and $\sqrt{s_{NN}}$ may be crucially important in order to
confirm this feature in the energy dependence of the important parameter
$\delta^{n}$. Furthermore there is a wide set of femtoscopic
measurements available at NICA-MPD with both the identical heavier
particles and the non-identical particle pairs. The $\Lambda
\Lambda$ correlations can be used for the search for exotic states in
QCD, for instance, $H$ dibaryons \cite{Greiner-PLB-219-199-1989}.
The non-identical particle correlations with kaons allow us to
obtain information about the space-time asymmetry of particle
emission, the final state interaction and some exotic objects like kaonic
atoms.

\section{Summary} \label{sec:5}

The absolute asymmetry is introduced in order to investigate the
evolution of CME in heavy ion collisions with initial energy.
Dependence on $\sqrt{s_{NN}}$ has been obtained for $A_{a}$ based
on the experimental correlators. The energy dependence of absolute
asymmetry for semicentral events in heavy ion collisions shows
sharp decreasing at $\sqrt{s_{NN}} < 19.6$ GeV, almost constant
behavior up to $\sqrt{s_{NN}} \simeq 200$ GeV with some decreasing
with further energy increasing. $A_{a}(\sqrt{s_{NN}})$ dependence
qualitatively corresponds to the energy dependence of the
Chern--Simons diffusion rate. $A_{a}(\sqrt{s_{NN}})$ indicates
possibly the onset of predominance of the hadronic states over the
phase of color degrees of freedom in the deconfinement state in the
matter created in heavy ion collisions with initial energies in the
domain $\sqrt{s_{NN}} \lesssim 11.5 - 19.6$ GeV. The energy dependence
is investigated for the main femtoscopic parameters deduced in the
framework of the Gauss approach. There is no dramatic change of
femtoscopic parameter values in $A+A$ with increasing 
$\sqrt{s_{NN}}$ in the domain of collision energies $\sqrt{s_{NN}}
\geq 5$ GeV. Some normalized HBT radii in $pp$ are significantly larger than those in $A+A$ collisions especially in
the TeV-region. The fit curves demonstrate qualitative agreement with
experimental $A+A$ data for $\lambda$ at all available collision
energies and for normalized HBT radii in the energy domain
$\sqrt{s_{NN}} \geq 5$ GeV.

The investigations of azimuthal and femtoscopic correlations at
intermediate energies with high statistics at NICA-MPD can provide
new important information for better understanding the structure of
the QCD vacuum as well as the relation between geometry and dynamic
features of creation of the secondary particle source. The
correlation measurements can be one of the focuses and significant
part of the NICA-MPD physics program.


\begin{thebibliography}{}
\bibitem{Okorokov-UJPA-1-1996-2013}
V.A. Okorokov, E.V. Sandrakova, Univ. J. Phys. Appl. \textbf{1},
196 (2013).
\bibitem{Lee-PRD-8-1226-1973}
T.D. Lee, Phys. Rev. D \textbf{8}, 1226 (1973); T.D. Lee, G.G.
Wick, Phys. Rev. D \textbf{9}, 2291 (1974).
\bibitem{Kharzeev-NPA-803-227-2008}
D.E. Kharzeev \emph{et al.}, Nucl. Phys. A \textbf{803}, 227
(2008).
\bibitem{Fukushima-PRD-78-074033-2008}
K. Fukushima \emph{et al.}, Phys. Rev. D \textbf{78}, 074033
(2008).
\bibitem{Okorokov-IJMPE-22-1350041-2013}
V.A. Okorokov, Int. J. Mod. Phys. E \textbf{22}, 1350041 (2013).
\bibitem{Voloshin-PRC-70-057901-2004}
S.A. Voloshin, Phys. Rev. C \textbf{70}, 057901 (2004).
\bibitem{Basar-arXiv-1202.2161-2012}
G. Basar, D.E. Kharzeev, arXiv: 1202.2161 [hep-ph].
\bibitem{Okorokov-arXiv-0908.2522-2009}
V.A. Okorokov, arXiv: 0908.2522 [nucl-th]; V.A Okorokov, Phys. At.
Nucl. Eng. \textbf{4}, 805 (2013).
\bibitem{Skokov-IJMPA-24-5925-2009}
V.V. Skokov \emph{et al.}, Int. J. Mod. Phys. A \textbf{24}, 5925
(2009); V. Voronyuk \emph{et al.}, Phys. Rev. C \textbf{83},
054911 (2011).
\bibitem{Cleymans-PRC-73-034905-2006}
J. Cleymans \emph{et al.}, Phys. Rev. C \textbf{73}, 034905
(2006).
\bibitem{Aggarwal-PRC-83-024901-2011}
M.M. Aggarwal \emph{et al.}, Phys. Rev. C \textbf{83}, 024901
(2011).
\bibitem{Kumar-NPA-904-905-256c-2013}
L. Kumar, Nucl. Phys. A \textbf{904-905}, 256c (2013).
\bibitem{Okorokov-arXiv-1312.4269}
V.A. Okorokov, arXiv: 1312.4269 [nucl-ex]. 2013; V.A. Okorokov,
Adv. High Energy Phys. \textbf{2015}, 790646 (2015).
\bibitem{Okorokov-arXiv-1504.08336}
V.A. Okorokov, arXiv: 1504.08336 [nucl-ex].
\bibitem{Valentin-book-1982}
L. Valentin, {\em Subatomic physics: nuclei and particles} V. {\bf
I} (Ermann, Paris, 1982); K.N. Mukhin, {\em Experimental nuclear
physics} V. {\bf I} (Energoatomizdat, Moscow, 1993).
\bibitem{Collins-book-1977}
P. Collins, {\em An introduction to Regge theory and high energy
physics} (Cambridge Univ. Press, Cambridge, 1977).
\bibitem{PDG-2014}
K.A. Olive \emph{et al.}, Chin. Phys. C\textbf{38}, 090001 (2014).
\bibitem{Okorokov-AHEP-2015-914170-2015}
V.A. Okorokov, Adv. High Energy Phys. \textbf{2015}, 914170
(2015).
\bibitem{Bourrely-arXiv-1202.3611}
C. Bourrely {\it et al.}, arXiv: 1202.3611 [hep-ph]. 2012.
\bibitem{Ilgenfritz-PLB-325-263-1994}
E.-M. Ilgenfritz, E.V. Shuryak, Phys. Lett. B \textbf{325}, 263
(1994).
\bibitem{Greiner-PLB-219-199-1989}
C. Greiner, B. M$\ddot{\mbox{u}}$ller, Phys. Lett. B \textbf{219},
199 (1989).

\end{thebibliography}
\end{document}